# Research with AI Agents

## How agentic systems are changing scientific work

Johannes Lotz[1] · Markus Wenzel[1,2]

[1] Fraunhofer Institute for Digital Medicine MEVIS, Lübeck, Germany
[2] Constructor University, Bremen, Germany

## Practical conclusions

- Agentic systems have arrived in research: literature search, data analysis, and programming can be delegated, and the researchers' role shifts toward steering and oversight.
- Agentic systems can be used most efficiently where the result can be checked against a predefined criterion. Systems built from validated modules whose combination can be traced can make validation easier.
- AI and agents are tools: whether a scientific text, a research result, or a product was created with their help says nothing, in itself, about its value. Humans remain the responsible authors: they decide what to delegate and how to review the results.
- We recommend institutionalizing the exchange of experience and best practices in the use of agentic systems.

**Keywords**
Artificial intelligence · Agentic systems · Automation

**Abstract**

**Background.** Agentic AI systems independently decompose tasks such as literature search, data analysis, and programming into subtasks, search the web, access databases, and execute code. This allows them to perform digital research tasks at high speed.
**Objectives.** Under what conditions does the use of agentic systems produce reliable efficiency gains, and which tasks remain with researchers?
**Materials and methods.** Summary of current studies on literature searches, data analysis, software development, and clinical decision support.
**Results.** For digital activities, work shifts from execution to steering and review. Efficiency gains are greatest when expected behavior can be formalized in advance and tested automatically. In complex agentic systems, recorded sequences of reasoning steps and tool calls are too extensive for human review. Furthermore, explanations generated by the model do not reliably reflect how an output was produced. One possible step toward more reliable systems is the validation of individual components. This limited reviewability extends beyond research itself; the review of scientific articles and grant proposals is also reaching capacity limits. In pathology, curated and annotated data, researchers' own analytical skills, and institutional exchange of experience are becoming increasingly important.
**Conclusions.** Researchers remain responsible for their results. They must determine what to delegate and how to review the results. The importance of a research question to patients, the field, and society cannot be fully assessed using formalized criteria and remains a matter of expert judgment. Agentic systems can free up time for this.

**AI agents** can search the literature, analyze data, and write code. They promise research in minutes instead of days. This gain does not come by itself. The faster texts, analyses, and proposals are produced, the more pressing the question of who reviews them, puts them into context, and takes responsibility for them. This article shows where agentic systems can take work off researchers' hands, and where the researchers' real work begins.

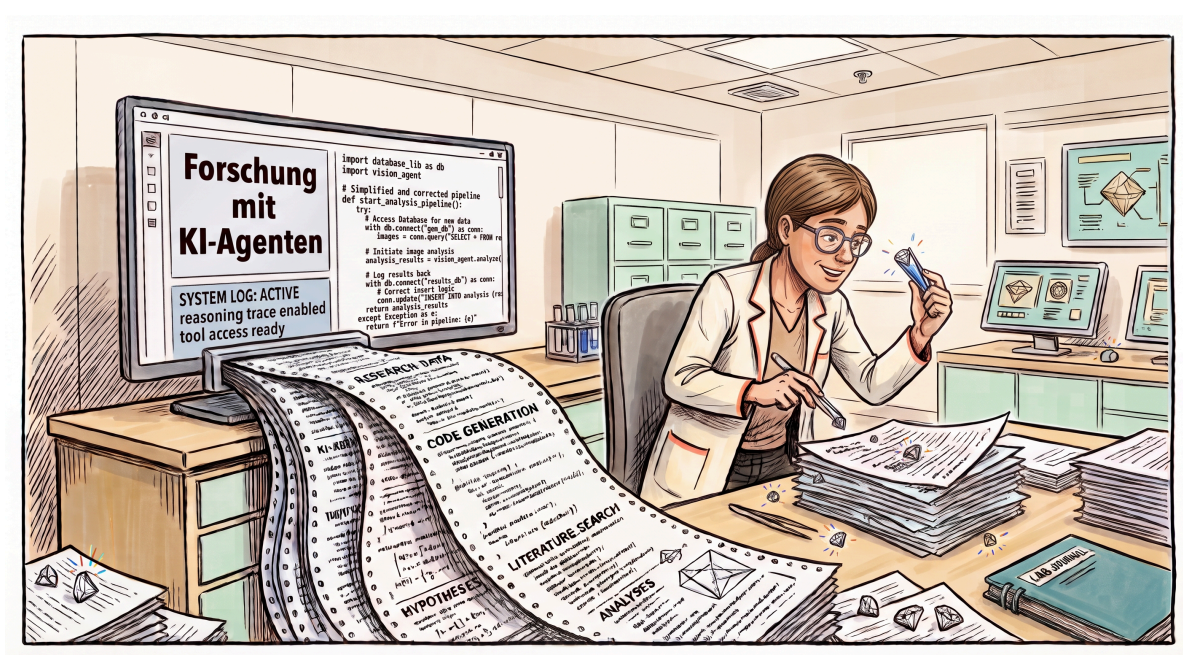

Figure 1: Agentic AI makes it easy and fast to generate texts and analyses. Researchers create scientific value by reviewing these outputs and putting them into context. (Own illustration; image created with the help of artificial intelligence (Google, Gemini 3 Pro Image, September 2026).)

## Agentic AI is changing research

Agentic AI systems extend language models with the ability to plan subtasks independently and to use external resources. These include web searches, database queries, and the execution of software and scripts [1]. For a more complex calculation, a system can generate and run a Python script instead of computing the result directly within the language model. This reduces hallucinations that can occur with purely model-internal computation [2].

Such systems are increasingly being used in research. Many institutions have issued recommendations for their use. Recurring requirements are transparency, data protection, and critical review of the results, as for instance in the guidelines of the German Research Foundation (DFG) [3].

In this article, we regard research as the search for open questions that enrich a scientific field with new insights and new connections between existing knowledge. This therefore includes identifying research gaps and judging which of them will lead to substantial advances. This judgment requires a deep understanding of the context. The problem that arises when research questions, of whatever complexity, can be delegated to an AI with a prompt is that the overview of the field and the "feel" for both the difficulty and the potential benefit of a question are lost. In a post on Mastodon [4], Terence Tao describes this as a "flattening of the difficulty profile" of a field and warns against it, since the flattening makes it ever harder to identify the few far-reaching research questions, and these still have to be identified by humans.

## The researchers' role is shifting

These new tools have the potential to fundamentally change the work of researchers in pathology and beyond. This is most advanced in working with text: at least 13 percent of the biomedical abstracts published in 2024 show statistically conspicuous word frequencies typical of AI ("delves", "crucial") [5]. The actual share has probably risen since then while becoming less obvious. Beyond merely drafting text, agentic systems can search for information on the internet or in the scientific literature, structure it, and incorporate it into their results.

This also shifts the error profile: language models without search access sometimes invent references that do not exist. This problem becomes rarer as soon as the system actually searches [6]. What remains are misattributions in which the source exists and fits the topic but does not support the specific claim; for current models with web search, this affected roughly a quarter to a half of the citations [7].

Agentic systems that independently write and run scripts for data analysis are already noticeably speeding up analysis. One example from pathology is the Cologne system SPARK [8]: based on a given task, the language model itself formulated hypotheses about relevant tissue parameters and translated them into executable code; from these parameters, microsatellite status, among other things, could then be predicted. The search space, however, was narrowly constrained, since the ideas could only be built from a few cell types and tissue compartments, which were supplied by a preprocessing step built on pathologist annotations.

The shift mainly affects digital processes; laboratory work can so far only be partially automated [9]. In the examples above, the researchers' role moves from execution to steering and oversight.

## Responsibility remains with humans

Automating individual tasks can considerably accelerate research and thus the discovery of new relationships; in drug development, rentosertib, a first AI-generated drug candidate, is approaching phase 3 testing after a positive phase 2 trial [10]. High efficiency, however, requires that high speed be matched by high quality. Speed has already been achieved: in programming, literature search, or writing, large amounts of code and text are produced in very little time. Quality, by contrast, cannot yet be reliably ensured. Humans should therefore critically read everything, which is hardly feasible in terms of time and demanding in terms of content. Anyone who receives 10 to 50 references after ten minutes of AI-assisted literature search has saved the preselection, while the reading remains. The preselection can be accelerated if the system provides a summary and verbatim supporting passages for each paper, but this does not replace reading the selected papers. This is compounded by a well-documented cognitive effect: people tend to confirm rather than question the decisions of automated systems ("automation complacency") [11].

To protect society from false medical findings, researchers must vouch for the correctness of their results. Santoni de Sio and Mecacci [12] argue that assigning responsibility creates the incentive to ensure this correctness. Since a computer system cannot bear responsibility, legal and moral responsibility must remain with humans ("culpability" and "moral accountability"). AI remains a tool whose use is decided by humans. For clinical pathology, this is already required by law: the EU AI Act demands human oversight for high-risk systems; AI outputs must be validated by a pathologist before clinical use [13].

However, only those who are actually able to check can bear responsibility. And only checking turns an AI output into a reliable result. The value of work produced with the help of AI systems therefore arises in its verification rather than in its generation. This is exactly where the current bottleneck lies, because humans can check the results of AI only to a very limited extent.

## Efficiency gains through automatable verification

The greatest efficiency gains arise where automated verification is possible. This applies first of all to software development itself, where compilers and automated tests ensure the formal correctness and the expected behavior of code. Human oversight remains necessary but shifts from reviewing individual lines of code to defining the verification criterion, for example to the question of whether a test written by the agent itself checks the right thing. In SPARK [8], every generated code snippet was tested for executability and runtime on ten test tiles from a TCGA slide; in case of errors, a code-review agent received the error message and could revise the code up to three times.

In this way, verification can be moved forward from the individual result to the tool. A laboratory analyzer is validated against reference standards and is not re-checked with every use thereafter. This validation presupposes that the task is fixed and that the validation data set represents it. Trustworthy AI systems therefore emerge first where the task is similarly well delimited; for clinical use, such validation on representative data sets is already mandatory [13]. An agent working on research questions, by contrast, receives a different task with every call. How such systems can be validated is an open question that goes beyond measuring average performance on benchmarks.

One approach is to validate individual modules: tools with a fixed task, such as an image analysis model or a database query, can be validated like an analyzer. What then remains unvalidated is the combination of these building blocks, that is, the agent's decision which module to call with which inputs. Criteria that would allow an automatic classification as right or wrong are lacking for it as soon as information from several databases or clinical systems is combined; it remains a matter of human review, which documented intermediate steps can shorten. Ferber et al. [14], for example, describe a decision-support system in oncology in which an agent calls image analysis models for histology and radiology, searches databases, and provides its answers with source references. Because each intermediate step is tied to a named source, the composition can be checked specifically instead of having to reconstruct the model's entire line of reasoning.

**The role of agentic AI in writing this article**

For the reasons given, this article was largely written by hand. At the same time, the authors greatly value the benefits of powerful AI models and test their added value and their limits in every task that comes up, including the writing of scientific contributions and articles such as this one.

The structure and argument emerged from discussions among colleagues, and the authors wrote the draft version of the text themselves (partly in bullet points); revision was partly carried out in dialogue with a language model, with every change in these sections checked by the authors and adopted under their responsibility. The position of this article is the authors' decision.

Agentic methods were used for the literature research, and this is where the real gain lay. A system with access to the internet, to PubMed, and to the full texts of the papers found searched and organized the literature, summarized individual papers to help us get up to speed (on request also as an audio version), supported its assessments with verbatim quotations from the full texts, and in a second pass searched specifically for supporting evidence and counterarguments for each section of the manuscript. Based on these summaries and quotations, the authors selected the literature. This allowed us to survey a field of considerable breadth in a way that would otherwise have been impossible in the same time. Errors occurred occasionally, such as (later corrected) incorrect or one-sided assessments of articles or incorrectly extracted author names.

The authors read all cited works themselves; the agentic research thus eased the selection of relevant literature without replacing the assessment of content.

## Complex systems elude automated verification

Both ways of easing the burden of review, automated verification against a predefined criterion and source-linked traceability, have limits. Automated verification requires that the expected behavior can be specified; targeted human review requires a manageable number of steps. Yet AI increasingly enables complex systems, in particular agents that respond dynamically to their context. Their behavior can at best be assessed empirically and in terms of their average performance. For more extensive audits, it has been proposed to store and evaluate so-called "reasoning traces" (their internal "thoughts"). But these traces are usually too long for human review.

Turpin et al. [15] showed that LLMs, when reasoning, not only demonstrably follow societal biases but also fail to document this in their "chain of thought"; further work confirms this behavior in current reasoning models as well [16].

Despite some success in eliciting even counterfactual reasoning from LLMs [17, 18], the currently prevailing scientific opinion largely denies language models the ability to make causal arguments as soon as they are expected to reproduce more than simple formal causal inferences in the sense of Pearl's do-calculus [19] [17]. The reasoning provided by the model is either a causal argument that could be trivially derived from the training data, or it is based on elementary calculations of causal effect sizes.

The arguments that, even more fundamentally, deny LLMs and AI the capacity for causal *understanding* [20] go beyond mechanistic understanding and have so far been addressed at best partially, not least because there is still no accepted operationalizable test for such understanding.

## What this means for research in pathology

For several years, the major international conferences have been confronted with floods of submissions whose text was written by AI; it has become apparent that AI-generated text cites incorrectly [21]. At the same time, conferences and journals are reaching the limits of their reviewing capacity [22].

The same applies to research funding: national and international funding programs are being flooded with proposals [23], which can be produced faster than ever with AI; evaluation can hardly cope with the volume without AI support. When AI in turn assesses AI, creativity and the setting of goals under human responsibility are undermined. A recent US analysis shows that proposals created with AI were closer in content to previously funded ideas [24]. Together with declining funding success rates, this development increases the need for alternatives to the proposal-based system of research funding.

In the hands of researchers who act ethically, AI can have a positive effect. It can protect the functioning of the funding system and perhaps even improve it. It can act as a twofold filter. First, researchers can compare their basic idea with the literature and thus better define the contribution they want to make. If this comparison means that proposals with little novelty are not submitted, peer review is relieved. Second, promising ideas can be worked out more comprehensively and more thoroughly. If this results in fewer but higher-quality proposals, funders can prioritize more strongly on scientific content rather than mainly checking technical feasibility. One example of this is the International Conference on Learning Representations (ICLR), which offers an AI-based pre-review at submission but leaves it to the authors how to deal with the feedback.

For pathology, the shift from execution to oversight is an opportunity. Vos et al. [25] describe how, in clinical practice, the role is shifting from pure diagnostics toward the integration and interpretation of data. The same holds for research, only earlier, because almost every step there is digital: researchers will less often carry out analyses themselves and more often determine what is analyzed, against what a result is to be measured, and whether the findings hold up.

Expertise thereby moves to the beginning of the process: when hypotheses and analyses are produced in minutes, data and annotations determine the value of the results. SPARK [8] could only operate within a search space that pathologists had annotated beforehand. Systems in neighboring disciplines also rely on reference data from pathology [14]. Building multicenter, well-annotated, and balanced datasets is therefore part of research rather than mere groundwork, and it increasingly determines the informative value of research results.

For this to succeed, the training of pathologists must preserve competencies that no longer seem to be needed in everyday work. Anyone who has never designed an analysis themselves cannot judge whether an agent is working correctly. Vos et al. [25] therefore name deskilling and automation bias as risks and propose corresponding learning objectives. Research institutions must themselves create the conditions for effective human oversight. To this end, they should institutionalize the exchange of experience and best practices in the use of agentic systems.

What remains is the question of which problems are worth solving. Agentic systems can identify research gaps and rank them by feasibility. Their significance for patients, the field, and society, however, cannot be determined by predefined criteria. Researchers must therefore decide what they delegate and how they review results; responsibility remains with them. The time gained belongs to the researchers.

## Corresponding address

Dr. rer. nat. Johannes Lotz
Fraunhofer Institute for Digital Medicine MEVIS
Maria-Goeppert-Str. 3, 23562 Lübeck, Germany
Phone +49 451 3101-6101
johannes.lotz@mevis.fraunhofer.de

## Compliance with ethical guidelines

**Conflict of interest.** The authors declare that they have no competing interests.

**Use of artificial intelligence.** Agentic language model systems with connected literature databases were used for literature research, checking supporting passages against the full texts, linguistic assistance with wording, and critical review of the argumentation and source attribution. All cited works were read by the authors themselves, and all statements and supporting passages were checked by them; responsibility for the content lies entirely with the authors. Figure 1 was created with generative AI; it is a schematic illustration without underlying data.

This article does not contain any studies with human participants or animals.

## Acknowledgments

The authors thank Daniel Budelmann, Stefan Heldmann, Nils Papenberg, Jan-Philip Redlich, Raphael Schäfer, and Nick Weiss for the discussions from which the position and argument of this article emerged, as well as Gabriele Lotz, Ole Schwen, and Till Nicke for critically reviewing the manuscript.

The project underlying this article was funded by the German Federal Ministry of Education and Research under grant number 03VP13371. The authors are responsible for the content of this publication.

*The references for both language versions are listed at the end of this document.*

Preprint

# Forschung mit Hilfe von KI-Agenten

## Wie agentische Systeme wissenschaftliches Arbeiten verändern

Johannes Lotz[1] · Markus Wenzel[1,2]

[1] Fraunhofer-Institut für digitale Medizin MEVIS, Lübeck
[2] Constructor University Bremen

### Fazit für die Praxis

- Agentische Systeme sind in der Forschung angekommen: Recherche, Datenauswertung und Programmierung lassen sich delegieren, die Aufgabe der Forschenden verschiebt sich zur Steuerung und Überwachung.
- Agentische Systeme können dort am effizientesten genutzt werden, wo sich das Ergebnis gegen ein vorab definiertes Kriterium prüfen lässt. Systeme aus validierten Modulen, deren Kombination sich nachvollziehen lässt, können die Validierung erleichtern.
- KI und Agenten sind Werkzeuge: Ob ein wissenschaftlicher Text, ein Forschungsergebnis oder ein Produkt mit ihrer Hilfe entstanden ist, sagt für sich genommen nichts über seinen Wert aus. Menschen bleiben die verantwortlichen Urheber: Sie legen fest, was sie delegieren und wie sie die Ergebnisse prüfen.
- Wir empfehlen, den Austausch über Erfahrungen und bewährte Verfahren beim Einsatz agentischer Systeme zu institutionalisieren.

**Schlüsselwörter**
Künstliche Intelligenz · Agentische Systeme · Automatisierung

**Zusammenfassung**

**Hintergrund.** Agentische KI-Systeme zerlegen Aufgaben wie Literaturrecherche, Datenauswertung und Programmierung selbstständig in Teilschritte, suchen im Web, greifen auf Datenbanken zu und führen Programmcode aus. Dadurch können sie digitale Forschungstätigkeiten in hoher Geschwindigkeit übernehmen.
**Fragestellung.** Unter welchen Bedingungen steigert der Einsatz agentischer Systeme die Effizienz belastbar, und welche Aufgaben bleiben bei den Forschenden?
**Material und Methoden.** Zusammenfassung aktueller Arbeiten zu Literaturrecherche, Datenanalyse, Softwareentwicklung und klinischer Entscheidungsunterstützung.
**Ergebnisse.** Bei digitalen Tätigkeiten verschiebt sich die Arbeit von der Ausführung zur Steuerung und Prüfung. Die größten Effizienzgewinne entstehen, wenn sich das erwartete Verhalten vorab formalisieren und automatisch testen lässt. Bei komplexen agentischen Systemen sind aufgezeichnete Abfolgen von Begründungsschritten und Werkzeugaufrufen zu umfangreich für ein menschliches Review. Zudem geben vom Modell erzeugte Begründungen die Entstehung einer Ausgabe nicht verlässlich wieder. Ein möglicher Schritt zu verlässlicheren Systemen ist die Validierung einzelner Komponenten. Diese begrenzte Prüfbarkeit betrifft nicht nur die Forschung an sich: Auch die Begutachtung wissenschaftlicher Artikel und Forschungsanträge stößt an Kapazitätsgrenzen. Für die Pathologie gewinnen kuratierte und annotierte Daten, eigene Analysekompetenz und institutioneller Erfahrungsaustausch an Bedeutung.
**Schlussfolgerungen.** Forschende bleiben für ihre Ergebnisse verantwortlich. Sie müssen festlegen, was sie delegieren und wie sie die Ergebnisse prüfen. Die Bedeutung einer Forschungsfrage für Patientinnen und Patienten, Fach und Gesellschaft lässt sich nicht vollständig anhand formalisierter Kriterien bewerten und bleibt eine fachliche Entscheidung. Agentische Systeme können dafür Zeit schaffen.

**KI-Agenten** können Literatur suchen, Daten auswerten und Code schreiben. Damit versprechen sie Forschung in Minuten statt Tagen. Der Gewinn entsteht aber nicht automatisch: Je schneller Texte, Analysen und Anträge entstehen, desto wichtiger wird die Frage, wer sie prüft, einordnet und verantwortet. Der Beitrag zeigt, wo agentische Systeme Forschende entlasten können und wo ihre Arbeit erst beginnt.

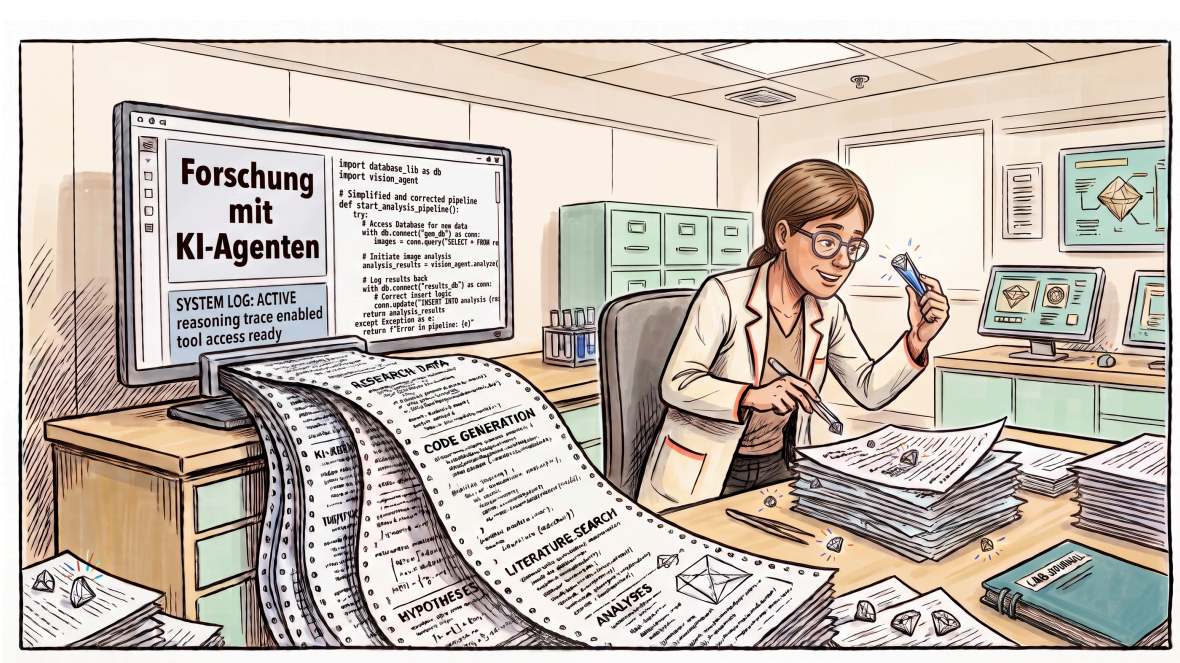

Abbildung 1: Mit agentischer KI lassen sich Texte und Auswertungen einfach und schnell erzeugen. Forschende schaffen wissenschaftlichen Wert, indem sie diese Ausgaben prüfen und einordnen. (Eigene Darstellung; Abbildung mithilfe Künstlicher Intelligenz erstellt (Google, Gemini 3 Pro Image, September 2026).)

## Agentische KI verändert die Forschung

Agentische KI-Systeme erweitern Sprachmodelle um die Fähigkeit, Teilaufgaben selbstständig zu planen und externe Ressourcen einzusetzen. Dazu zählen Websuchen, Datenbankabfragen sowie die Ausführung von Software und Skripten [1]. Bei komplexeren Rechenaufgaben kann ein System beispielsweise ein Python-Skript erzeugen und ausführen, statt das Ergebnis unmittelbar im Sprachmodell zu berechnen. Dies reduziert Halluzinationen, die bei rein modellinterner Berechnung auftreten können [2].

Solche Systeme werden zunehmend in der Forschung eingesetzt. Viele Institutionen haben Empfehlungen für ihre Nutzung erlassen. Wiederkehrende Anforderungen sind Transparenz, Datenschutz und die kritische Prüfung der Ergebnisse, etwa in den Leitlinien der DFG [3].

Forscherisches Tun betrachten wir in diesem Artikel als die Suche nach offenen Fragen, die ein wissenschaftliches Gebiet um neue Erkenntnisse und neue Verbindungen zwischen bekanntem Wissen bereichern. Darunter fällt also die Identifikation von Forschungslücken und die Beurteilung, welche davon zu signifikanten Neuerungen führen. Diese Beurteilung setzt ein tiefes Verstehen des Kontexts voraus. Das Problem, das entsteht, wenn Forschungsfragen – egal welcher Komplexität – durch einen Prompt an eine KI delegiert werden können, ist, dass die Übersicht über das Gebiet und das „Gespür" sowohl für die Schwierigkeit als auch den potentiellen Nutzen einer Frage verloren gehen. Terence Tao bezeichnet das in einem Beitrag auf Mastodon [4] als „Glättung des Schwierigkeitsprofils" eines Arbeitsfelds und warnt davor, denn die Glättung mache die Identifikation der wenigen weitreichenden Forschungsfragen immer schwerer – und sie müssen noch immer von Menschen identifiziert werden.

## Die Aufgabe der Forschenden verschiebt sich

Diese neuen Werkzeuge haben das Potential, die Arbeit von Forschenden in der Pathologie und darüber hinaus grundlegend zu verändern. Am weitesten fortgeschritten ist dies bei der Arbeit an Texten: In mindestens 13 Prozent der 2024 erschienenen biomedizinischen Abstracts finden sich KI-typische statistisch auffällige Worthäufungen („delves", „crucial") [5]. Der tatsächliche Anteil dürfte seither gestiegen und zugleich weniger offensichtlich geworden sein. Über das bloße Formulieren hinaus können agentische Systeme Informationen aus dem Internet oder der wissenschaftlichen Literatur suchen, strukturieren und in ihre Ergebnisse einbeziehen.

Damit verschiebt sich auch das Fehlerbild: Sprachmodelle ohne Suchzugriff erfinden mitunter Literaturstellen, die nicht existieren. Dieses Problem wird seltener, sobald das System tatsächlich sucht [6]. Es bleiben Fehleinordnungen, bei denen die Quelle existiert und zum Thema passt, die konkrete Aussage aber nicht deckt; bei aktuellen Modellen mit Websuche betraf das rund ein Viertel bis die Hälfte der Belege [7].

Agentische Systeme, die eigenständig Skripte zur Datenauswertung schreiben und ausführen, beschleunigen die Auswertung bereits jetzt spürbar. Ein Beispiel aus der Pathologie ist das Kölner System SPARK [8]: Auf Basis einer vorgegebenen Aufgabenstellung formulierte hier das Sprachmodell selbst Hypothesen zu relevanten Gewebeparametern und setzte sie in lauffähigen Code um; aus diesen Parametern ließ sich anschließend unter anderem der Mikrosatellitenstatus vorhersagen. Der Suchraum war allerdings eng begrenzt – die Ideen konnten nur aus wenigen Zelltypen und Gewebekompartimenten bestehen, die eine von Pathologinnen und Pathologen annotierte Vorverarbeitung lieferte.

Die Verschiebung betrifft vor allem digitale Prozesse; Labortätigkeiten lassen sich bislang nur teilweise automatisieren [9]. In den genannten Beispielen verlagert sich die Aufgabe der Forschenden von der Ausführung zur Steuerung und Überwachung.

## Die Verantwortung bleibt beim Menschen

Die Automatisierung einzelner Tätigkeiten kann die Forschung und damit die Entdeckung neuer Zusammenhänge erheblich beschleunigen; in der Wirkstoffentwicklung steht mit Rentosertib nach einer positiven Phase-2-Studie ein erster KI-generierter Wirkstoffkandidat vor der Phase-3-Prüfung [10]. Hohe Effizienz setzt allerdings voraus, dass hohe Geschwindigkeit mit hoher Qualität einhergeht. Die Geschwindigkeit ist bereits erreicht: Beim Programmieren, bei der Literaturrecherche oder beim Verfassen von Texten entstehen große Mengen an Code und Text in kürzester Zeit. Die Qualität lässt sich dagegen noch nicht verlässlich sichern. Menschen sollten daher alles kritisch mitlesen, was zeitlich kaum möglich und inhaltlich aufwendig ist: Wer nach zehn Minuten KI-gestützter Literatursuche 10 bis 50 Referenzen erhält, hat nicht das Lesen gespart, sondern die Vorauswahl. Diese lässt sich beschleunigen, wenn das System zu jeder Arbeit eine Zusammenfassung und wörtliche Belegstellen liefert – die Lektüre der ausgewählten Arbeiten ersetzt das nicht. Hinzu kommt ein gut dokumentierter kognitiver Effekt: Menschen neigen dazu, Entscheidungen automatischer Systeme eher zu bestätigen als zu hinterfragen („automation complacency") [11].

Um die Gesellschaft vor falschen medizinischen Erkenntnissen zu schützen, müssen Forschende für die Richtigkeit ihrer Ergebnisse einstehen. Santoni de Sio und Mecacci [12] argumentieren, dass die Zuschreibung von Verantwortung den Anreiz schafft, diese Richtigkeit sicherzustellen. Da ein Computersystem keine Verantwortung übernehmen kann, müssen rechtliche und moralische Verantwortung beim Menschen bleiben („culpability" und „moral accountability"). Die KI bleibt ein Werkzeug, über dessen Verwendung Menschen entscheiden. Für die klinische Pathologie ist dies bereits gesetzlich vorgeschrieben: Der EU AI Act verlangt bei Hochrisikosystemen menschliche Aufsicht; KI-Ausgaben müssen vor der klinischen Verwendung durch eine Pathologin oder einen Pathologen validiert werden [13].

Verantwortung tragen kann allerdings nur, wer tatsächlich in der Lage ist zu prüfen. Und erst die Prüfung macht aus einer KI-Ausgabe ein belastbares Ergebnis: Der Wert einer mit Hilfe von KI-Systemen erzeugten Arbeit entsteht nicht bei der Erzeugung, sondern bei der Kontrolle. Genau hier liegt aktuell der Flaschenhals, denn Menschen können die Ergebnisse der KI nur sehr begrenzt kontrollieren.

## Effizienzgewinn durch automatisierbare Prüfung

Die größten Effizienzgewinne entstehen dort, wo eine automatisierte Prüfung möglich ist. Das gilt zunächst für die Softwareentwicklung selbst, in der Compiler und automatische Tests die formale Korrektheit und das erwartete Verhalten des Codes absichern. Die menschliche Aufsicht bleibt dabei nötig, verschiebt sich aber vom Review einzelner Codezeilen zur Definition des Prüfkriteriums, etwa zur Frage, ob ein vom Agenten selbst geschriebener Test das Richtige prüft. In SPARK [8] wurde jeder generierte Codeschnipsel an zehn Testkacheln eines TCGA-Schnitts auf Lauffähigkeit und Laufzeit geprüft; bei Fehlern erhielt ein Code-Review-Agent die Fehlermeldung und konnte den Code bis zu dreimal überarbeiten.

Die Prüfung lässt sich so vom einzelnen Ergebnis auf das Werkzeug vorverlagern. Ein Laboranalysegerät wird gegen Referenzstandards validiert und danach nicht mehr bei jeder Verwendung geprüft. Diese Validierung setzt voraus, dass die Aufgabe feststeht und der Validierungsdatensatz sie repräsentiert. Vertrauenswürdige KI-Systeme entstehen deshalb zunächst dort, wo die Aufgabe ähnlich klar umrissen ist; für die klinische Anwendung ist eine solche Validierung auf repräsentativen Datensätzen bereits vorgeschrieben [13]. Ein Agent, der Forschungsfragen bearbeitet, erhält dagegen bei jedem Aufruf eine andere Aufgabe. Wie sich solche Systeme validieren lassen, ist eine offene Frage, die über die Messung der durchschnittlichen Leistung auf Benchmarks hinausgeht.

Ein Ansatz ist die Validierung einzelner Module: Werkzeuge mit fester Aufgabe, etwa ein Bildanalysemodell oder eine Datenbankabfrage, lassen sich wie ein Analysegerät validieren. Unvalidiert bleibt dann die Kombination der Bausteine, also die Entscheidung des Agenten, welches Modul er mit welchen Eingaben aufruft. Für sie fehlen Kriterien, die eine automatische Einstufung als richtig oder falsch erlauben, sobald Informationen aus mehreren Datenbanken oder klinischen Systemen zusammenfließen; sie bleibt Sache menschlicher Prüfung, die durch dokumentierte Zwischenschritte verkürzt werden kann. So beschreiben Ferber et al. [14] ein System zur Entscheidungsunterstützung in der Onkologie, in dem ein Agent Bildanalysemodelle für Histologie und Radiologie aufruft, in Datenbanken recherchiert und seine Antworten mit Quellenangaben versieht. Weil jeder Zwischenschritt an eine benannte Quelle gebunden ist, lässt sich gezielt die Zusammensetzung prüfen, statt die gesamte Argumentation aus dem Modell heraus nachvollziehen zu müssen.

**Die Rolle agentischer KI beim Verfassen dieses Artikels**

Dieser Artikel ist aus den angeführten Gründen zu großen Teilen von Hand geschrieben. Gleichzeitig schätzen die Autoren den Nutzen starker KI-Modelle sehr und erproben ihren Mehrwert und ihre Grenzen in jeder sich stellenden Aufgabe – so auch bei der Erstellung wissenschaftlicher Beiträge und Artikel wie dem vorliegenden.

Gliederung und Argumentation sind aus Diskussionen unter Kolleginnen und Kollegen entstanden, die Konzeptversion des Textes haben die Autoren selbst (teilweise stichpunktartig) formuliert; die Überarbeitung erfolgte teilweise im Dialog mit einem Sprachmodell, wobei in diesen Abschnitten jede Änderung von den Autoren geprüft und verantwortlich übernommen wurde. Die Position des Beitrags ist eine Entscheidung der Autoren.

Agentisch gearbeitet wurde bei der Recherche, und hier lag der eigentliche Gewinn. Ein System mit Zugriff auf das Internet, auf PubMed und auf die Volltexte der gefundenen Arbeiten suchte und ordnete die Literatur, fasste einzelne Arbeiten zur Einarbeitung zusammen, auf Wunsch auch als Hörfassung, belegte seine Einordnung mit wörtlichen Zitaten aus den Volltexten und suchte in einem zweiten Durchgang zu jedem Abschnitt gezielt nach Belegen und nach Widerreden. Anhand dieser Zusammenfassungen und Zitate trafen die Autoren die Auswahl der Literatur. Dies ließ uns ein Feld beträchtlicher Breite sichten, wie es in der gleichen Zeit anders unmöglich gewesen wäre. Fehler traten dabei gelegentlich auf, etwa (später korrigierte) fehlerhafte oder einseitige Bewertungen von Artikeln oder falsch extrahierte Autorennamen.

Die Autoren haben alle zitierten Arbeiten selbst gelesen; die agentische Recherche ersetzte damit nicht die inhaltliche Prüfung, sondern erleichterte die Auswahl relevanter Literatur.

## Komplexe Systeme entziehen sich der automatisierten Prüfung

Beide Formen der Entlastung – die automatisierte Prüfung gegen ein vorab definiertes Kriterium und die quellengebundene Nachvollziehbarkeit – haben Grenzen. Eine automatisierte Prüfung setzt voraus, dass sich das erwartete Verhalten festschreiben lässt; eine gezielte menschliche Prüfung erfordert eine überschaubare Zahl von Schritten. KI ermöglicht jedoch zunehmend komplexe Systeme, insbesondere Agenten, die dynamisch auf ihren Kontext reagieren. Ihr Verhalten lässt sich allenfalls empirisch und anhand ihrer durchschnittlichen Leistung bewerten. Für weitergehende Audits wird vorgeschlagen, sogenannte „Reasoning Traces“ (ihre internen „Gedanken“) zu speichern und auszuwerten. Doch diese Gedanken-Spuren sind meist zu lang für ein menschliches Review.

Turpin et al. [15] konnten zeigen, dass LLMs beim Schlussfolgern nicht nur offensichtlich gesellschaftlichen Biases folgen, sondern dies zusätzlich in ihrer „Chain of Thought“ nicht dokumentieren; weitere Arbeiten bestätigen dieses Verhalten auch bei aktuellen Reasoning-Modellen [16].

Trotz einiger Erfolge, LLMs auch kontrafaktische Begründungen abzuringen [17, 18], spricht die aktuell vorherrschende wissenschaftliche Meinung Sprachmodellen die Fähigkeit zu kausalen Argumenten weitgehend ab, sobald sie mehr als einfache formale Kausalschlüsse im Sinne von Pearls Do-Kalkül [19] reproduzieren sollen [17]. Die vom Modell gelieferte Begründung ist entweder ein kausales Argument, das trivial aus den Trainingsdaten abgeleitet werden konnte, oder sie basiert auf elementaren Berechnungen von kausalen Effektstärken.

Die Argumente, die LLMs und KI noch grundsätzlicher die Fähigkeit zum kausalen *Verstehen* absprechen [20], gehen über das mechanistische Verständnis hinaus und sind bis heute allenfalls teilweise adressiert, was auch daran liegt, dass es noch keinen anerkannten operationalisierbaren Test für dieses Verstehen gibt.

## Was das für die forschende Pathologie heißt

Die wichtigsten internationalen Konferenzen werden seit einigen Jahren mit Fluten von Einreichungen konfrontiert, deren Text durch KI verfasst wurde; dabei zeigte sich, dass KI falsch zitiert [21]. Gleichzeitig stoßen Konferenzen und Journale an die Grenzen ihrer Begutachtungs-Kapazität [22].

Das Gleiche gilt für die Forschungsförderung: Nationale und internationale Förderformate werden von Anträgen überflutet [23], die sich mit KI schneller denn je erstellen lassen; die Evaluierung kann der Menge kaum noch ohne KI-Unterstützung begegnen. Wenn KI wiederum KI beurteilt, werden Kreativität und menschlich verantwortetes Setzen von Zielen ausgehebelt. Eine aktuelle US-amerikanische Analyse zeigt dabei, dass mit KI erstellte Anträge inhaltlich näher an bereits geförderten Ideen lagen [24]. Zusammen mit sinkenden Förderquoten verstärkt diese Entwicklung den Bedarf an Alternativen zum antragsbasierten System der Forschungsförderung.

KI kann in den Händen ethisch handelnder Forschender positive Wirkung entfalten und die Funktion des Systems absichern, vielleicht sogar verbessern. Dabei kann sie als zweifacher Filter wirken. Erstens können Forschende ihre grundsätzliche Idee mit der Literatur abgleichen und damit besser definieren, welchen möglichen Beitrag sie leisten wollen. Führt dieser Abgleich dazu, dass Anträge mit geringer Erfindungshöhe nicht eingereicht werden, entlastet dies das Peer Review. Zweitens lassen sich aussichtsreiche Ideen umfassender und fundierter ausarbeiten. Werden dadurch weniger, aber hochwertigere Anträge eingereicht, können Fördermittelgeber stärker inhaltlich priorisieren, statt vor allem die technische Machbarkeit zu prüfen. Einen ähnlichen Ansatz verfolgt die International Conference on Learning Representations (ICLR): Sie bietet bei der Einreichung ein KI-basiertes Vorabreview an, überlässt den Umgang mit der Rückmeldung jedoch den Autorinnen und Autoren.

Für die Pathologie ist die Verschiebung von der Ausführung zur Überwachung eine Chance. Vos et al. [25] beschreiben für die Klinik, dass sich die Rolle von der reinen Diagnostik hin zur Integration und Interpretation von Daten wandelt. In der Forschung gilt dasselbe, nur früher, weil dort fast jeder Arbeitsschritt digital ist: Wer forscht, wird seltener selbst auswerten und häufiger festlegen, was ausgewertet wird, woran ein Ergebnis zu messen ist und ob das Gefundene trägt.

Die Expertise verlagert sich dabei an den Anfang des Prozesses: Wenn Hypothesen und Analysen in Minuten entstehen, bestimmen Daten und Annotationen den Wert der Ergebnisse. SPARK [8] konnte nur innerhalb eines Suchraums arbeiten, den Pathologinnen und Pathologen zuvor annotiert hatten. Auch Systeme in Nachbarfächern beruhen auf pathologischen Referenzdaten [14]. Der Aufbau multizentrischer, gut annotierter und ausgewogener Datensätze ist damit weniger denn je bloße Vorarbeit. Er ist ein Teil der Forschung und bestimmt zunehmend die Aussagekraft ihrer Ergebnisse.

Damit das gelingt, muss die Ausbildung Kompetenzen erhalten, die im Alltag scheinbar nicht mehr gebraucht werden. Wer nie selbst eine Auswertung konzipiert hat, kann nicht beurteilen, ob ein Agent korrekt arbeitet. Vos et al. [25] benennen deshalb Deskilling und Automatisierungsbias als Risiken und schlagen entsprechende Lernziele vor. Forschungseinrichtungen müssen die Voraussetzungen für eine wirksame menschliche Aufsicht selbst schaffen. Dazu sollten sie den Austausch über Erfahrungen und bewährte Verfahren beim Einsatz agentischer Systeme institutionalisieren.

Was bleibt, ist die Frage, welche Probleme es wert sind, gelöst zu werden. Agentische Systeme können Forschungslücken identifizieren und nach Machbarkeit ordnen. Ihre Bedeutung für Patientinnen und Patienten, Fach und Gesellschaft lässt sich jedoch nicht anhand eines vordefinierten Maßstabs bestimmen. Forschende müssen daher festlegen, was sie delegieren und wie sie Ergebnisse prüfen; die Verantwortung bleibt bei ihnen. Die dabei gewonnene Zeit gehört den Forschenden.

## Korrespondenzadresse

Dr. rer. nat. Johannes Lotz
Fraunhofer-Institut für digitale Medizin MEVIS
Maria-Goeppert-Str. 3, 23562 Lübeck
Tel. +49 451 3101-6101
johannes.lotz@mevis.fraunhofer.de

## Einhaltung ethischer Richtlinien

**Interessenkonflikt.** Die Autoren geben an, dass kein Interessenkonflikt besteht.

**Einsatz Künstlicher Intelligenz.** Für Literaturrecherche, Prüfung von Belegstellen gegen die Volltexte, sprachliche Formulierungshilfen sowie kritische Durchsicht von Argumentation und Quellenzuordnung wurden agentische Sprachmodell-Systeme mit angeschlossenen Literaturdatenbanken eingesetzt. Alle zitierten Arbeiten wurden von den Autoren selbst gelesen und alle Aussagen und Belegstellen von ihnen geprüft; die Verantwortung für den Inhalt liegt vollständig bei den Autoren. Abbildung 1 wurde mit generativer KI erzeugt; sie ist eine schematische Illustration ohne zugrunde liegende Daten.

Dieser Beitrag beinhaltet keine Studien an Menschen oder Tieren.

## Danksagung

Die Autoren danken Daniel Budelmann, Stefan Heldmann, Nils Papenberg, Jan-Philip Redlich, Raphael Schäfer und Nick Weiss für die Diskussionen, aus denen die Position und Argumentation dieses Beitrags hervorgegangen sind, sowie Gabriele Lotz, Ole Schwen und Till Nicke für die kritische Durchsicht des Manuskripts.

Das diesem Artikel zugrunde liegende Vorhaben wurde mit Mitteln des Bundesministeriums für Bildung und Forschung unter dem Förderkennzeichen 03VP13371 gefördert. Die Verantwortung für den Inhalt dieser Veröffentlichung liegt bei der Autorin / beim Autor.

## Literaturverzeichnis / References